\documentstyle[preprint,aps,epsfig]{revtex}

\def\be{\begin{equation}} 
\def\ee{\end{equation}}
\def\tbfm{{\mathord{\buildrel{\lower3pt\hbox{$\scriptscriptstyle\leftrightarrow$}}\over {\bf m}}}}
\def\tbfM{{\mathord{\buildrel{\lower3pt\hbox{$\scriptscriptstyle\leftrightarrow$}}\over {\bf M}}}}
\def\tm{{\mathord{\buildrel{\lower3pt\hbox{$\scriptscriptstyle\leftrightarrow$}}\over  m}}}
\def\tmu{{\mathord{\buildrel{\lower3pt\hbox{$\scriptscriptstyle\leftrightarrow$}}\over \mu}}}
\def\tI{{\mathord{\buildrel{\lower3pt\hbox{$\scriptscriptstyle\leftrightarrow$}}\over I}}}
\def\tg{{\mathord{\buildrel{\lower3pt\hbox{$\scriptscriptstyle\leftrightarrow$}}\over g}}}
\def\tbfe{{\mathord{\buildrel{\lower3pt\hbox{$\scriptscriptstyle\leftrightarrow$}}\over {\bf e}}}}
\def\tbfB{{\mathord{\buildrel{\lower3pt\hbox{$\scriptscriptstyle\leftrightarrow$}}\over {\bf B}}}}
\def\tbfH{{\mathord{\buildrel{\lower3pt\hbox{$\scriptscriptstyle\leftrightarrow$}}\over {\bf H}}}}
\def\tH{{\mathord{\buildrel{\lower3pt\hbox{$\scriptscriptstyle\leftrightarrow$}}\over H}}}
\def\t0{{\mathord{\buildrel{\lower3pt\hbox{$\scriptscriptstyle\leftrightarrow$}}\over 0}}}
\def\tbfrho{{\mathord{\buildrel{\lower3pt\hbox{$\scriptscriptstyle\leftrightarrow$}}\over {\bf \rho}}}}
\def\tbfP{{\mathord{\buildrel{\lower3pt\hbox{$\scriptscriptstyle\leftrightarrow$}}\over {\bf P}}}}
\def\tP{{\mathord{\buildrel{\lower3pt\hbox{$\scriptscriptstyle\leftrightarrow$}}\over  P}}}
\def\tb{{\mathord{\buildrel{\lower3pt\hbox{$\scriptscriptstyle\leftrightarrow$}}\over  b}}}
\def\tU{{\mathord{\buildrel{\lower3pt\hbox{$\scriptscriptstyle\leftrightarrow$}}\over  U}}}
\def\trho{{\mathord{\buildrel{\lower3pt\hbox{$\scriptscriptstyle\leftrightarrow$}}\over \rho}}}
\def\tGamma{{\mathord{\buildrel{\lower3pt\hbox{$\scriptscriptstyle\leftrightarrow$}}\over \Gamma}}}
\def\txi{{\mathord{\buildrel{\lower3pt\hbox{$\scriptscriptstyle\leftrightarrow$}}\over \xi}}}
\def\tbeta{{\mathord{\buildrel{\lower3pt\hbox{$\scriptscriptstyle\leftrightarrow$}}\over \beta}}}
\def\talpha{{\mathord{\buildrel{\lower3pt\hbox{$\scriptscriptstyle\leftrightarrow$}}\over \alpha}}}

\begin{document}
\draft
                 
\title{Statistical-mechanical theory of the overall magnetic properties \\ of  mesocrystals}
\author{J. P. Huang}


\address{Department of Physics, The Chinese University of Hong Kong, Shatin, NT, Hong Kong, and \\
Max Planck Institute for Polymer Research, Ackermannweg 10, 55128 Mainz, Germany}

\maketitle

\begin{abstract}

The mesocrystal showing both electrorheological and magnetorheological effects is called  electro-magnetorheological (EMR)  solids. Prediction of the overall magnetic properties of the EMR solids is a challenging task due to the coexistence of the uniaxially anisotropic behavior and structural transition as well as long-range interaction between the suspended particles. To consider the uniaxial anisotropy effect, we present an anisotropic Kirkwood-Fr\"{o}hlich equation for calculating the effective permeabilities by adopting an explicit characteristic spheroid rather than a characteristic sphere used in the derivation of the usual  Kirkwood-Fr\"{o}hlich equation. Further,  by applying an Ewald-Kornfeld formulation we are able to investigate the effective permeability by including the structural transition and long-range interaction explicitly. Our theory can reduce to the usual  Kirkwood-Fr\"{o}hlich equation and Onsager equation naturally. To this end, the numerical simulation shows the validity of monitoring the structure of EMR solids by detecting their effective permeabilities.

\end{abstract}

\pacs{PACS: 75.30.Gw,  64.70.Kb, 75.50.Kj }





\newpage

\section{introduction}

In 1998 and 1999 a new mesocrystal was reported~\cite{Tao98,Sheng99} which combines both electrorheological (ER) or magnetorheological (MR) effects. This sort of mesocrystal is also called electro- and magneto-rheological (EMR)  solids. In fact, ER~\cite{Win49} and MR~\cite{book} fluids are generally particle suspensions in which the particles have large electric polarizability or magnetic permeability. In the application of an external electric or magnetic field, the suspended particles can form body-centered tetragonal (bct) mesocrystallities, namely ER or MR solids~\cite{bct}.
The EMR solid shows very interesting properties when the applied electric field $E$ (in  $z$ axis) is perpendicular to the magnetic field $H$ (in  $x$ or $y$ axis). In detail, in  case of dominate electric field or dominate magnetic field, EMR solids have thick columns in the dominate field direction. These columns have a bct lattice as the ideal structure, too. Recently,  a novel structural transition in EMR solids was theoretically~\cite{Tao98} and experimentally~\cite{Sheng99} observed from bct to face-centered cubic (fcc) lattice in the presence of crossed electric and magnetic fields as the ratio between the magnetic field and electric field exceeds a minimum value.

Understanding the magnetic properties of EMR solids is critical to the  design of EMR-fluid-based devices. Also, these magnetic properties may provide valuable insight into the character of the microstructure responsible for their field-dependent rheology as well as  models of the EMR effect.
Since for EMR solids the uniaxial anisotropy occurs naturally, the magnetic properties in longitudinal fields (L) should be different from those in tranverse fields (T). Furthermore, the  structural transition can affect the effective magnetic properties, and the longe-range interaction between the particles (lattice effect) should be expected to play an important role as well.
Thus, prediction of the overall magnetic properties of EMR solids is indeed a challenging task.


To calculate the effective permeability of EMR solids, the existing methods for cubic arrays of spheres~\cite{Doyle}  or for a suspension containing a dense array of particles~\cite{Keller} can not be used directly. 
Recently, one developed a theory of homogenization to study the effective permeability of MR solids with a periodic microstructure~\cite{Simon}. In this paper, we shall present a statistical-mechanical   theory, in order to calculate the effective permeability of the EMR solids.

This paper is organized as follows. In Sec.~II, by developing the  Kirkwood-Fr\"{o}hlich equation and using the Ewald-Kornfeld formulation,   we present a statistical-mechanical theory for  the effective permeability of the EMR solids, and the  numerical results are given as well. This paper ends with a discussion and conclusion in Sec.~III.

\section{Formalism and numerical results}

\subsection{Contribution of permanent magnetic moments}

For an EMR solid, its effective permeability  $\tmu_e$ 
is uniaxially anisotropic due to the application of external fields.
In detail,  the transverse component $ \mu_e^{(T)}$ (in $x$ or $y$ axis)   differs from the longitudinal component  $\mu_e^{(L)}$ (in $z$ axis). In this connection, the effective permeability $\tmu_e$ should possess a tensorial form like
\be
\tmu_e =  \left( \begin{array}{ccc}
          \mu_e^{(T)} & 0 & 0 \\
          0 &  \mu_e^{(T)}  & 0 \\
          0 & 0 & \mu_e^{(L)}
          \end{array}\right).\label{mue}
\ee
Since all the permeable particles of the EMR solid have a permanent magnetic dipole moment ${\bf m}$, it becomes more difficult to derive the expression for $\tmu_e$. For this purpose,   Kirkwood~\cite{Kirkwood} and Fr\"{o}hlich~\cite{Fro}  introduced a continuum with permeability $\tmu_{e0}$ which arises from induced magnetization only.
Based on it, we shall  derive the effective permeability  $\tmu_e$  of the EMR solid consisting of  permeable particles with a permanent magnetic moment  ${\bf m}$. Let us start by seeing each particle with ${\bf m}$ to have a new tensorial moment $\tbfm'$,
\be
\tbfm' = \frac{\tmu_{e0}+2\mu_2\tI}{3\mu_2\tI}{\bf m}\label{moment}
\ee
and to be embedded in a new host (introduced continuum) of $\tmu_{e0}$, where $\mu_2$ denotes the permeability of the nonmagnetic carrier fluid, and $\tI$ a unit matrix. The denominators in Eq.~(\ref{moment}) and in the following equations should be interpreted as inverse matrices.   In this model each particle is replaced by a point dipole $\tbfm'$ having the same non-electrostatic interactions with the other point dipoles as the particles had, while the magnetizability of the particles can be imagined to be smeared out to form a continuum with permeability $\tmu_{e0}$, which will be derived in Sec.~(\ref{induced}). 
 Next, to include the anisotropic feature, we take a characteristic spheroid of volume $V$, which contains $n$ particles.  In doing so, the  particles in the spheroid will be treated explicitly by taking into account the contribution of the particle interaction to the effective permeability. In principle, the approximation in this method can be made as small as necessary by taking $n$ sufficiently large. Here we should remark that for discussing isotropic cases  Kirkwood~\cite{Kirkwood} and Fr\"{o}hlich~\cite{Fro}  used a characteristic sphere. As a matter of fact, no matter for a sphere or a spheroid, each of them should reflect the physical properties of the whole suspension. For instance, the number density inside the sphere or spheroid should be identical to that of the whole system under consideration. In this regard, for the present EMR solid a characteristic sphere is far from being satisfactory, and  a characteristic spheroid can be used instead so that the uniaxially anisotropic behavior of the suspension may be considered more physically. We shall show that the explicit spheroidal shape of choice can be determined exactly, see Eq.~(\ref{relation}) below.

All statistical-mechanical theories of the permeability start from  
\be
{\bf B} - \mu_2 {\bf H} = 4 \pi {\bf \rho}_t,\label{de1}
\ee
where ${\bf B}$ and ${\bf H}$ denotes the magnetic induction and  Maxwell field in the material outside the spheroid, respectively.  By definition, we  write for the magnetization density ${\bf \rho}_t$ as
\be
{\bf \rho}_t V = \langle {\bf M}_t\rangle ,
\label{de2}
\ee
where $\langle {\bf M}_t\rangle$ stands for the average total magnetic moment of the spheroid. Here and below $\langle\cdots\rangle$ stands for a statistical mechanical average, e.g., 
$$
\langle {\bf M}_t\rangle = \frac{\int {\rm d}X {\bf M}_t \exp^{-U/kT}}{\int {\rm d}X \exp^{-U/kT}}.
$$
In this expression, $X$ stands for the set of position and orientation 
variables of all particles. Here $U$ is the 
energy related to the dipoles in the spheroid, and it consists of three parts:
the energy of the dipoles in the external field, the magnetostatic interaction energy of the dipoles, and the non-magnetostatic interaction energy between the dipoles which are 
responsible for the short-range correlation between orientations and 
positions of the dipoles.

Then, the tensorial effective permeability $\tmu_e$ of the whole system can be defined as
\be
{\bf B}\tI = \tmu_e{\bf H} .\label{de3}
\ee
In view of Eqs.~(\ref{de2})~and~(\ref{de3}), we take one step forward to rewrite Eq.~(\ref{de1}) as
\be
(\tmu_e - \mu_2\tI){\bf H} = \frac{4\pi\tI}{V} \langle {\bf M}_t \rangle .
\ee
Since $ \langle {\bf M}_t \rangle$ has the same direction as ${\bf H}$, it suffices to calculate the average component of ${\bf M}_t$ in the direction of ${\bf H}$, thus we have 
\be
(\tmu_e - \mu_2 \tI){\bf H}  = \frac{4\pi\tI}{V} \langle{\bf M}_t\cdot {\hat {\bf e}} \rangle.
\ee
Here ${\hat {\bf e}}$ is the  unit vector in the direction of the field.

In general, ${\bf \rho}_t$ and $\langle {\bf M}_t \rangle$ contain  also terms in higher powers of ${\bf H}$. Thus, $(1/4\pi)(\tmu_{e0}-\mu_2\tI)H$ is the first term in a series development of ${\bf \rho}_i\tI$ (induced magnetization) in powers of $H$, and must be set equal to the term linear in $H$ of the series development of $ \langle {\bf M}_t\cdot  {\hat {\bf e}} \rangle\tI$ in a Taylor series. So, we obtain
\be
\tmu_e-\mu_2\tI = \frac{4\pi\tI}{V} \left (\frac{\partial \langle {\bf M}_t\cdot  {\hat {\bf e}}  \rangle}{\partial H}          \right )_{H = 0}.
\ee   
Owing to  $\langle {\bf M}_t\cdot  {\hat {\bf e}} \rangle = V ({\bf \rho}_{i} + {\bf \rho}_{o} )$ and ${\bf \rho}_i\tI =(1/4\pi) (\tmu_{e0}-\mu_2\tI) {\bf H}$, we have
\be
\tmu_e-\tmu_{e0} = 4\pi\tI \left (\frac{\partial \rho_o}{\partial H} \right )_{H = 0},
\ee
where $\rho_o$ stands for the orientational magnetization arising from the permanent magnetic moments. Rewriting with the external field $H_0$ instead of the Maxwell field $H$ as the independent variable we obtain
\be
\tmu_e-\tmu_{e0} = \frac{4\pi\tI}{V} \left (\frac{\partial H_0}{\partial H}  \right )_{H = 0} \left (\frac{\partial \langle {\bf M}_o\cdot  {\hat {\bf e}} \rangle }{\partial H_0} \right )_{H_0 = 0},\label{C0}
\ee
where $\langle{\bf M}_o\cdot  {\hat {\bf e}}\rangle = V\rho_o$.
In this case, the external field acting on the spheroid is
\be
{\bf H}_0\tI =   \left( \begin{array}{ccc}
          \frac{\mu_e^{(T)}}{\mu_e^{(T)}+(\mu_{e0}^{(T)}-\mu_e^{(T)})g^{(T)} }  & 0 & 0 \\
          0 &  \frac{\mu_e^{(T)}}{\mu_e^{(T)}+(\mu_{e0}^{(T)}-\mu_e^{(T)})g^{(T)} }   & 0 \\
          0 & 0 &  \frac{\mu_e^{(L)}}{\mu_e^{(L)}+(\mu_{e0}^{(L)}-\mu_e^{(L)})g^{(L)} }
          \end{array}\right) {\bf H} \equiv \txi {\bf H} ,\label{C1}
\ee
where the tensorial depolarization factor $\tg$
represents the spheroid shape. In fact, the degree of field-induced anisotropy of the system is determined by how $\tg$ deviates from $1/3\tI$ (isotropic limit). It is worth noting that $\tg$ will be determined explicitly [see Eq.~(\ref{relation})], and that its components satisfy a sum rule $2g^{(T)}+g^{(L)}=1$~\cite{Landau}.

Starting from 
\be
\left (\frac{\partial \langle {\bf M}_o\cdot  {\hat {\bf e}} \rangle }{\partial H_0} \right )_{H_0 = 0} = -\frac{1}{kT}\left\langle {\bf M}_o\cdot  {\hat {\bf e}} \frac{\partial U}{\partial H_0} \right\rangle_{H_0 = 0},
\ee
eventually we have
\be
\left (\frac{\partial \langle {\bf M}_o\cdot  {\hat {\bf e}} \rangle }{\partial H_0} \right )_{H_0 = 0} = \frac{1}{3kT}\left\langle {\bf M}_o^2 \right\rangle_{H_0 = 0}.\label{C2}
\ee

If we use  a tensorial Kirkwood correlation factor $\tbeta$,
  then we  obtain 
\be
\tI\langle {\bf M}_o^2 \rangle_{H_0 = 0} = n\tm'^2\tbeta.\label{C3}
\ee  
In view of Eqs.~(\ref{C1}),~(\ref{C2})~and~(\ref{C3}), Eq.~(\ref{C0}) can be rewritten as
\be
\tmu_e-\tmu_{e0} = \frac{4\pi N}{3kT}\tm'^2\txi \tbeta ,\label{Huang}
\ee
where $N$ denotes the number density of the particles.
 For an isotropic system, namely $\tg = 1/3\tI$,  Eq.~(\ref{Huang}) reduces to the usual Kirkwood-Fr\"{o}hlich equation~\cite{Fro} which works for permeable particles with a permanent magnetic moment. If  $\tg = 1/3\tI$,
  $\tbeta = \tI$  and $\tmu_{e0} = \tI$, Eq.~(\ref{Huang}) reduces to the  Onsager equation~\cite{Onsager}  which treats non-permeable particles with a permanent magnetic moment embedded in vacuum. However, it is worth noting that in the derivation of the Onsager equation only one  particle  is considered in the characteristic sphere. That is, there is no more correlations    between the particle orientations than can be accounted for with the help of the continuum method, thus yielding $\tbeta =\tI$.

\subsection{Contribution of induced magnetic moments}\label{induced}

Now we are in a position to derive the induced-magnetization-related permeabibility $\tmu_{e0}$ by performing an Ewald-Kornfeld formulation~\cite{Ewald,Lo01} so that the structural transition and long-range interaction can be  taken into account explicitly. The ground state of the EMR  solid is a bct (body-centered tetragonal) lattice, which can be regarded as a
tetragonal lattice, plus a basis of two particles each of which is fixed with an induced point magnetic dipole at its center. One of the two particles is located at a corner and
the other one at the body center of the tetragonal unit cell. Its
lattice constants are denoted by $a_3=q\ell$ and $a_1(=a_2)=\ell q^{-1/2}$ along the $z-$ and $x-(y-)$ axes, respectively. In this case, the
uniaxial anisotropic axis is directed along $z$ axis.
 As $q$ varies, the volume of the unit cell keeps unchanged, i.e. $V_c=\ell^3 .$ Thus, the degree of anisotropy of the tetragonal lattice is measured by how $q$ is deviated from unity. In particular, $q=0.87358$, $1$ and $2^{1/3}$ represents the bct, bcc (body-centered cubic) and fcc lattice, respectively.

When one applies an  external magnetic field ${\bf H}_0$ along $x$ axis, the induced dipole moment
${\bf P}$ are perpendicular to the uniaxial anisotropic axis.
Then, the local field
${\bf H}_{lc}$
  at the lattice point ${\bf R}={\bf 0}$
can be determined. Let us take the transverse component as an example, and resort to the Ewald-Kornfeld formulation~\cite{Ewald,Lo01} to calculate the local field $H_{lc}$ such that
\be
H_{lc} = P\sum_{j=1}^2\sum_{\vec{R}\ne \vec{0}}[-\gamma_1(R_j)+x_j^2q^2\gamma_2(R_j)]-\frac{4\pi P}{V_c}\sum_{\vec{G}\ne \vec{0}}\Pi(\vec{G})\frac{G_x^2}{G^2}\exp(\frac{-G^2}{4\eta^2})+\frac{4P\eta^3}{3\sqrt{\pi}}\label{Ex}.
\ee
In this equation, $\gamma_1$ and $\gamma_2$ are two coefficients, given by

\begin{eqnarray}
\gamma_1(r)&=&\frac{{\rm erfc}(\eta r)}{r^3}+\frac{2\eta}{\sqrt{\pi}r^2}\exp(-\eta^2r^2),\nonumber\\
\gamma_2(r)&=&\frac{3{\rm erfc}(\eta r)}{r^5}+(\frac{4\eta^3}{\sqrt{\pi}r^2}+\frac{6\eta}{\sqrt{\pi}r^4})\exp(-\eta^2r^2),\nonumber
\end{eqnarray}
where ${\rm erfc}(\eta r)$ is the complementary error function, and $\eta$ an adjustable parameter making the summation
converge rapidly. In Eq.~(\ref{Ex}),
$R$ and $G$ denote the lattice vector and the reciprocal lattice vector,  respectively,
\begin{eqnarray}
\vec{R}&=&\ell (q^{-1/2}l\hat{x}+q^{-1/2}m\hat{y}+qn\hat{z}),\nonumber\\
\vec{G}&=&\frac{2\pi}{\ell}(q^{1/2}u\hat{x}+q^{1/2}v\hat{y}+q^{-1}w\hat{z}),\nonumber
\end{eqnarray}
where $l,m,n,u,v,w$ are integers. In addition,  $x_j$ and $R_j$  of Eq.~(\ref{Ex})  are given by, 
$$
x_j=l-\frac{j-1}{2},\,\, R_j=|\vec{R}-\frac{j-1}{2}(a\hat{x}+a\hat{y}+c\hat{z})|,
$$
and the structure factor $\Pi(\vec{G})=1+\exp[i(u+v+w)/\pi]$.

So far, let us define a local field factor $\talpha$,
\be
\talpha = \frac{3V_c}{4\pi}\frac{H_{lc}}{P}\tI.
\ee
 It is worth remarking  that $\talpha$ is a function of a single variable, namely degree of anisotropy $q .$ Also, there is a sum rule $2\alpha^{(T)}+\alpha^{(L)}=3$~\cite{PRE1}. As $q=1$, $\talpha = \tI$  just represents the isotropic limit. Next, we take one step forward to rewrite the well-known Maxwell-Garnett theory for  isotropic suspensions as~\cite{Lo01,PRE1,Maxwell} 
\be
\frac{\tmu_{e0}-\mu_2\tI}{\talpha\tmu_{e0}+(3\tI-\talpha)\mu_2} = f\frac{\mu_1-\mu_2}{\mu_1+2\mu_2}\tI,\label{AMGA}
\ee
where $\mu_1$ stands for the permeability of the particles. This is a developed Maxwell-Garnett theory for  uniaxially anisotropic suspensions~\cite{Lo01,PRE1}. Then it is not difficult to see that the depolarization factor $\tg$ [Eq.~(\ref{C1})] characterizing the shape of the characteristic spheroid of choice is determined by
\be
\tg = \frac{1}{3}\talpha. \label{relation}
\ee
The substitution of $\tmu_{e0}$ [obtained from Eq.~(\ref{AMGA})] into Eq.~(\ref{Huang}) leads to  $\tmu_{e}$ as a result.

\subsection{Numerical results}

Let us do some numerical simulations. Figure 1 displays  $\mu_e^{(T)}$ and  $\mu_e^{(L)}$ as a function of $q$. For this figure, we used the Onsager consideration (i.e., assuming the characteristic spheroid contains  only one particle, $\tbeta = \tI$), with a focus on the anisotropic effect. As $q=1$, this system is in the isotropic limit,  yielding $\mu_e^{(T)}=\mu_e^{(L)}$. Thus, in Fig.~1 the two points at $q=1$ are overlapped.   It is found that the structural transition of the EMR solid (measured by the variation of $q$) can cause $\tmu_e$ to change. 
To some extent, the numerical simulations show the validity of monitoring the structure of EMR solids by detecting their effective permeabilities.


\section{Discussion and conclusion}

Here some comments are in order. An approximation expression for the Kirkwood correlation factor $\tbeta$ can be obtained by taking only nearest-neighbors interactions into account~\cite{Bot}. In this case, the characteristic spheroid may be shrunk to contain only the $i$-th particle and all the nearest neighbors. It is apparent that $\beta^{(L)}$ or $\beta^{(T)}$ will be different from 1 when there is correlation between the orientations of neighboring particles. When the particles tend to direct themselves with parallel permanent  magnetic moments, $\beta^{(L)}$ or $\beta^{(T)}$ will be larger than 1. When the particles prefer an ordering with anti-parallel moments, $\beta^{(L)}$ or $\beta^{(T)}$ will be smaller than 1. As the EMR solid  is subjected to the external magnetic field, all the particles can easily direct themselves with parallel  permanent magnetic moments. In this connection, $\beta^{(L)}$ or $\beta^{(T)}$ should be larger than $1$, or could approximately equal to $1+N_c$ where $N_c$ denotes the number of the closest neighboring particles. Nevertheless, once $N_c>0$, the correlation between the nearest particles is included approximately, and this  (no figures shown here)  does not affect the present numerical result on the anisotropic effect as $\beta^{(L)} = \beta^{(T)} = 1$, i.e., $N_c=0$. In particular, as $N_c=4$ (i.e., for a BCT lattice), we obtain $\mu_e^{(T)}=17.40$ and $\mu_e^{(L)}=18.69$. They both are larger than those of $N_c=0$ due to the correlation, as expected.

The Bergman-Milton spectral representation (BMSR)~\cite{Bergman}  is an effective method for calculating the effective dielectric constant of a two-phase composite, and has been successfully applied in electrorheological fluids~\cite{MaPRL96}, in order to discuss the frequency-dependent complex dielectric constant. Alternatively, the BMSR should  be expected to work for EMR solids, and a favorable comparison between the BMSR and the Ewald summation technique used in this work is expected.

To sum up, the aim of the present work is to develop a statistical-mechanical  theory in order to calculate the effective permeability of a new mesocrystal (EMR solid). This theory allows one to study the overall magnetic properties of EMR solids, by taking into account the anisotropy and structural transition effects and the long-range interaction between the suspended particles. Our theory is  expected to be of value in computer simulations of  magnetic/dielectric properties of EMR fluids.

\section*{Acknowledgments}

This work was in part supported by the Alexander von Humboldt Foundation, Germany.  The author acknowledges Professor K. W. Yu's  fruitful discussions.

 \newpage
\begin{figure}[h]
\caption{(Color online). (a) $\mu_e^{(T)}$ and (b) $\mu_e^{(L)}$ as a function of $q$. Dot-dashed lines: bct ($q=0.87358$), bcc ($q=1$), and fcc ($q=2^{1/3}$).   Parameters: $\mu_1=2000$, $\mu_2=1$, $m=5.8\times 10^{-11}\,$emu,  $f=0.2$, $N=4.2\times 10^6\,$cm$^{-3}$,  and $T=298\,$K. Solid lines are a guide for the eye. }
\end{figure}

\newpage
\centerline{\epsfig{file=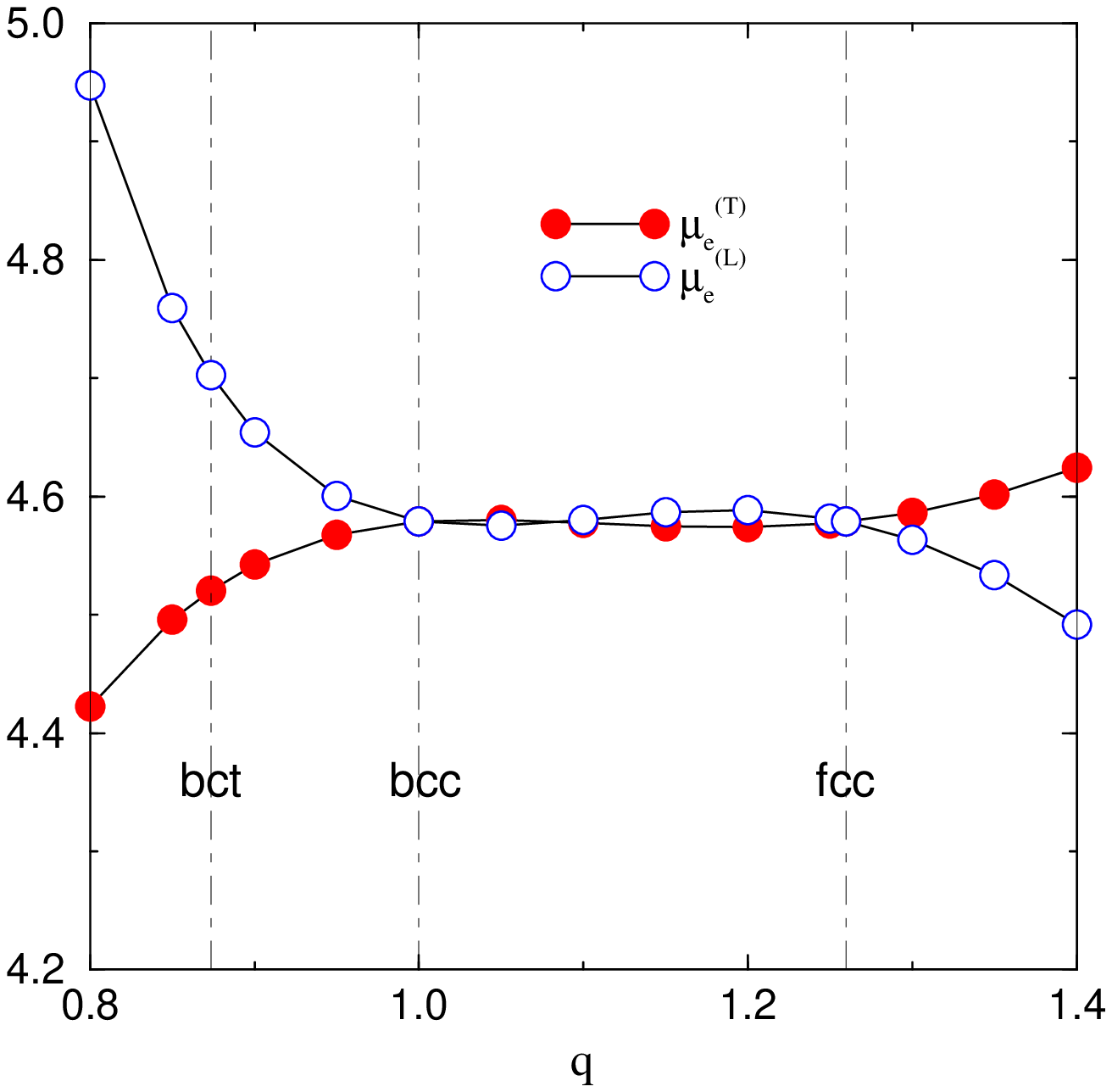,width=300pt}}
\centerline{Fig.~1/Huang}


\newpage

\begin{references}




\bibitem{Tao98} R. Tao and Q. Jiang, Phys. Rev. E {\bf 57}, 5761 (1998). 


\bibitem{Sheng99} W.  Wen, N. Wang, H.  Ma, Z.  Lin, W. Y.  Tam, C. T.  Chan, and P. Sheng, Phys. Rev. Lett. {\bf 82}, 4248 (1999).


\bibitem{Win49} For example, see  W. M. Winslow, J. Appl. Phys. {\bf 20}, 1137 (1949); T. C. Halsey, Science {\bf 258}, 761 (1992);  {\it Electrorheological Fluids}, edited by R. Tao (World Scientific, Singapore, 1992);  H. J. H. Clercx and G. Bossis, Phys. Rev. E {\bf 48}, 2721 (1993); D. J. Klingenberg, MRS Bull. {\bf 23}, 30 (1998);  U. Dassanayake, S. Fraden, and A. V. Blaaderen, J. Chem. Phys. {\bf 112}, 3851 (2000).


\bibitem{book} For example, see V. I. Kordonsky and Z. P. Shulman, in {\it Electrorheological Fluids}, edited by J. D. Carlson, A. F. Sprecher, and H. Conrad (Technomic Publishing, Lancaster, Basel, 1991), pp. 437-444; S. Cutillas and G. Bossis, Europhys. Lett. {\bf 40}, 465 (1997); J. M. Ginder, MRS BULL. {\bf 23}, 26 (1998);   S. Melle and J. E. Martin, J. Chem. Phys. {\bf 118}, 9875 (2003); Lord Corporation's homepage: http://www.mrfluid.com.

\bibitem{bct} R. Tao and J. M. Sun, Phys. Rev. Lett. {\bf 67}, 398 (1991); {\it ibid.}, Phys. Rev. A {\bf 44}, R6181 (1991); G. Bossis, H. Clerx, Y. Grasselli, and E. Lemaice, in {\it Electrorheological Fluids}, editted by R. Tao and G. D. Roy (World Scientific, Singapore, 1994), p. 153; L. Zhou, W. Wen, and P. Sheng, Phys. Rev. Lett. {\bf 81}, 1509 (1998).

\bibitem{Doyle} W. T. Doyle, J. Appl. Phys. {\bf 49}, 795 (1978).
\bibitem{Keller} J. B. Keller, J. Appl. Phys. {\bf 34}, 991 (1963).
\bibitem{Simon} T. M. Simon, F. Reitich, M. R. Jolly, K. Ito, H. T. Banks, Math. Comput. Model. {\bf 33}, 273 (2001).

\bibitem{Kirkwood} J. G. Kirkwood, J. Chem. Phys. {\bf 7}, 911 (1939).
\bibitem{Fro} 
H. Fr\"{o}hlich, {\em Theory of dielectrics} (Oxford University 
Press, London 1958).

\bibitem{Landau} L. D. Landau, E. M. Lifshitz, and L. P. Pitaevskii, {\it Electrodynamics of Continuous Media}, 2nd ed. (Pergamon Press, New York, 1984), Chap.~II.

\bibitem{Onsager} L. Onsager, J. Am. Chem. Soc. {\bf 58}, 1486 (1936).

\bibitem{Ewald} P. P. Ewald, Ann. Phys. (Leipzig) {\bf 64}, 253 (1921); H. Kornfeld, Z. Phys. {\bf 22}, 27 (1924).

\bibitem{Lo01} C. K. Lo and K. W. Yu, Phys. Rev. E {\bf 64} (2001) 031501.



\bibitem{PRE1} J. P. Huang, J. T. K. Wan, C. K. Lo, and K. W. Yu, Phys. Rev. E {\bf 64}, R061505 (2001).



\bibitem{Maxwell} J. C. M. Garnett, Philos. Trans. T. Soc. London {\bf 203}, 385, 1904; {\it ibid.,} {\bf 205}, 237, 1906.




\bibitem{Bot} 
C.\,J.\,F. B\"{o}ttcher, {\em Theory of electric polarization}, Vol.\,1, 
(Elsevier, Amsterdam, 1993).

\bibitem{Bergman} D. J. Bergman, in {\it Solid State Physics} Vol. 46, edited by H. Ehrenreich and D. Turnbull (Academic Press, New York, 1992), p.147; G. W. Milton, Appl. Phys. A {\bf 26}, 1207 (1981); G. W. Milton, J. Appl. Phys. {\bf 52}, 5286 (1980).

\bibitem{MaPRL96} H. Ma, W. Wen, W. Y. Tam, and P. Sheng, Phys. Rev. Lett. 77, 2499 (1996); W. Wen, H. Ma, W. Y. Tam, and P. Sheng, Phys. Rev. E 55, R1294 (1997).

\end{references}
\end{document}